\documentclass[12pt,preprint]{aastex}

\usepackage{natbib}
\bibpunct{(}{)}{,}{a}{}{,}
\usepackage{array}
\usepackage{tabularx}
\shorttitle{SOLAR MASS EVOLUTIONARY SEQUENCE WITH \textit{KEPLER} DATA}
\shortauthors{SILVA AGUIRRE ET AL.}

\begin{document}

\title{Constructing a one-solar-mass evolutionary sequence using asteroseismic data from \textit{Kepler}}

\author{V. Silva Aguirre\altaffilmark{1}, W.~J. Chaplin\altaffilmark{2}, J. Ballot\altaffilmark{3,4}, S. Basu\altaffilmark{5}, T.~R.Bedding\altaffilmark{6}, A.~M. Serenelli\altaffilmark{7}, G.~A. Verner\altaffilmark{2,8}, A. Miglio\altaffilmark{2}, M.~J.~P.~F.~G. Monteiro\altaffilmark{9}, A. Weiss\altaffilmark{1}, T. Appourchaux\altaffilmark{10}, A.  Bonanno\altaffilmark{11}, A.~M. Broomhall\altaffilmark{2}, H. Bruntt\altaffilmark{12}, T.~L. Campante\altaffilmark{9,12}, L. Casagrande\altaffilmark{1}, E. Corsaro\altaffilmark{11}, Y. Elsworth\altaffilmark{2}, R.~A. Garc\'\i a\altaffilmark{13}, P. Gaulme\altaffilmark{10}, R. Handberg\altaffilmark{12}, S. Hekker\altaffilmark{2,14}, D. Huber\altaffilmark{6}, C. Karoff\altaffilmark{12}, S. Mathur\altaffilmark{15}, B. Mosser\altaffilmark{16}, D. Salabert\altaffilmark{17}, R. Sch\"{o}nrich\altaffilmark{1}, S.~G. Sousa\altaffilmark{9}, D. Stello\altaffilmark{6}, T.~R. White\altaffilmark{6,18}, J. Christensen-Dalsgaard\altaffilmark{12}, R.~L. Gilliland\altaffilmark{19}, S.~D Kawaler\altaffilmark{20}, H. Kjeldsen\altaffilmark{12}, G. Houdek\altaffilmark{21}, T.~S. Metcalfe\altaffilmark{15}, J. Molenda-{\.Z}akowicz\altaffilmark{22}, M.~J. Thompson\altaffilmark{15}, D.~A. Caldwell\altaffilmark{23}, J.~L. Christiansen\altaffilmark{23}, \& B. Wohler\altaffilmark{24}}
\altaffiltext{1}{Max Planck Institute for Astrophysics, Karl-Schwarzschild-Str. 1, 85748, Garching bei M\"{u}nchen, Germany}
\altaffiltext{2}{School of Physics and Astronomy, University of Birmingham, Birmingham, B15 2TT, UK}
\altaffiltext{3}{Institut de Recherche en Astrophysique et Plan\'etologie, CNRS, 14 avenue Edouard Belin, 31400 Toulouse, France}
\altaffiltext{4}{Universit\'e de Toulouse, UPS-OMP, IRAP, Toulouse, France}
\altaffiltext{5}{Department of Astronomy, Yale University, P.O. Box 208101, New Haven, CT 06520-8101, USA}
\altaffiltext{6}{Sydney Institute for Astronomy (SIfA), School of Physics, University of Sydney, NSW 2006, Australia}
\altaffiltext{7}{Instituto de Ciencias del Espacio (CSIC-IEEC), Facultad de  Ci\`encies, Campus UAB, 08193 Bellaterra, Spain}
\altaffiltext{8}{Astronomy Unit, Queen Mary University of London, London, E1 4NS, UK}
\altaffiltext{9}{Centro de Astrof\'{\i}sica and Faculdade de Ci\^encias, Universidade do Porto, Rua das Estrelas, 4150-762 Porto, Portugal}
\altaffiltext{10}{Institut d'Astrophysique Spatiale UniversitŽ Paris 11 - CNRS (UMR8617) Batiment 121, F-91405 ORSAY Cedex}
\altaffiltext{11}{INAF -- Osservatorio Astrofisico di Catania, Via S. Sofia 78, I-95123 Catania}
\altaffiltext{12}{Department of Physics and Astronomy, Aarhus University, DK-8000 Aarhus C, Denmark}
\altaffiltext{13}{Laboratoire AIM, CEA/DSM-CNRS-Universit\'e Paris Diderot; IRFU/SAp, Centre de Saclay, 91191 Gif-sur-Yvette Cedex, France}
\altaffiltext{14}{Astronomical Institute 'Anton Pannekoek', University of Amsterdam, Science Park 904, 1098 XH Amsterdam, The Netherlands}
\altaffiltext{15}{High Altitude Observatory, NCAR, P.O. Box 3000, Boulder, CO 80307, USA}
\altaffiltext{16}{LESIA, CNRS, Universit\'e Pierre et Marie Curie, Universit\'e Denis Diderot, Observatoire de Paris, 92195 Meudon cedex, France}
\altaffiltext{17}{Universit\'e de Nice Sophia-Antipolis, CNRS, Observatoire de la C\^ote d'Azur, BP 4229, 06304 Nice Cedex 4, France}
\altaffiltext{18}{Australian Astronomical Observatory, PO Box 296, Epping NSW 1710, Australia}
\altaffiltext{19}{Space Telescope Science Institute, 3700 San Martin Drive, Baltimore, MD 21218, USA}
\altaffiltext{20}{Department of Physics and Astronomy, Iowa State University, Ames, IA, USA 50014}
\altaffiltext{21}{Institute of Astronomy, University of Vienna, A-1180 Vienna, Austria}
\altaffiltext{22}{Instytut Astronomiczny Uniwersytetu Wroc{\l}awskiego, ul.\ Kopernika 11, 51-622 Wroc{\l}aw, Poland}
\altaffiltext{23}{SETI Institute/NASA Ames Research Center, Moffett Field, CA 94035, USA}
\altaffiltext{24}{Orbital Sciences Corporation/NASA Ames Research Center, Moffett Field, CA 94035, USA}

\begin{abstract}
Asteroseismology of solar-type stars has entered a new era of large surveys with the success of the NASA \textit{Kepler} mission, which is providing exquisite data on oscillations of stars across the Hertzprung-Russell (HR) diagram. From the time-series photometry, the two seismic parameters that can be most readily extracted are the large frequency separation ($\Delta\nu$) and the frequency of maximum oscillation power ($\nu_\mathrm{max}$). After the survey phase, these quantities are available for hundreds of solar-type stars. By scaling from solar values, we use these two asteroseismic observables to identify for the first time an evolutionary sequence of 1-M$_\odot$ field stars, without the need for further information from stellar models. Comparison of our determinations with the few available spectroscopic results shows an excellent level of agreement. We discuss the potential of the method for differential analysis throughout the main-sequence evolution, and the possibility of detecting twins of very well-known stars.
\end{abstract}

\section{Introduction}\label{intro}
The high-precision photometric observations obtained by the NASA \textit{Kepler} Mission \citep{wb09a} have led to a dramatic increase in the number of main-sequence and subgiant stars with detected oscillations \citep{wc10,gill10a}. Data on hundreds of these stars are now available to test our knowledge of solar-like oscillations and their dependence on stellar parameters. Moreover, extremely precise information can be extracted on fundamental stellar properties and compared to population synthesis models of our galaxy \citep[e.g.][]{wc11a}.

The large ensemble of stars with detected oscillations allows us to select cohorts of targets sharing one or more common properties. By performing comparative studies on these stars, it should be possible to suppress the dependence of the modeling results on the shared characteristic, opening an exciting possibility for further constraining the internal physical processes. This is the basis for \textit{differential asteroseismology}: a detailed seismic comparison of the inner structures of stars showing some similar property.

To exploit this technique, one must be able to identify stars with common characteristics. Comparing the quantities extracted from conventional observations such as spectroscopy is not sufficient to fully constrain stellar properties, and evolutionary models must be employed to determine e.g. masses and radii. Asteroseismology allows us to go one step further by providing a tool to find pairs or groups of stars with similar characteristics.

In this Letter, we use global seismic properties of solar-type targets observed by \textit{Kepler} to construct an observational sequence of field stars with masses similar to that of the Sun, in a model-independent way.
\section{Extracting mass and radius from asteroseismic data} \label{method}
Two asteroseismic parameters can be readily extracted from the $p$-mode oscillation spectrum without the need for individual frequency determinations. One is the large frequency separation, $\Delta\nu$, which denotes the frequency difference between modes of the same degree and consecutive radial order. The other is the frequency of maximum oscillation power, $\nu_\mathrm{max}$. It has been shown that, to very good approximation, $\Delta\nu$ scales as the square root of the mean density \citep[e.g.][]{ru86}, while $\nu_\mathrm{max}$ is related to the dynamical timescale of the atmosphere, as represented by the acoustic cut-off frequency \citep[e.g.][]{tb91,kb95}. These two quantities are tightly correlated over a wide range of values, and follow scaling relations from the accurately known solar parameters \citep[e.g.][]{sh09,ds09a,sh11a}. These scaling relations can be written as 
\begin{equation}\label{eqn:mass} 
\frac{M}{M_\odot} \simeq \left(\frac{\nu_{\mathrm{max}}}{\nu_{\mathrm{max},\odot}}\right)^{3} \left(\frac{\Delta\nu}{\Delta\nu_\odot}\right)^{-4}\left(\frac{T_\mathrm{eff}}{T_{\mathrm{eff},\odot}}\right)^{3/2}, 
\end{equation}
\begin{equation}\label{eqn:rad} 
\frac{R}{R_\odot} \simeq \left(\frac{\nu_{\mathrm{max}}}{\nu_{\mathrm{max},\odot}}\right) \left(\frac{\Delta\nu}{\Delta\nu_\odot}\right)^{-2}\left(\frac{T_\mathrm{eff}}{T_{\mathrm{eff},\odot}}\right)^{1/2}, 
\end{equation}
where $T_\mathrm{eff}$ is the effective temperature of the star, and $\Delta\nu_\odot = 135.1\,\rm \mu Hz$, $\nu_{\mathrm{max},\odot}= 3150\,\rm \mu Hz$, $T_{\mathrm{eff},\odot}= 5777\,\rm K$ are the observed values in the Sun \citep[e.g.][]{wc11b}. From these two equations it is evident that, provided we have a measurement of $T_\mathrm{eff}$, the global seismic observables give a determination of stellar mass and radius for a given star that is independent of evolutionary models. This is the so-called \textit{direct method}, and it has been used to constrain the fundamental properties of red giant stars and distinguish different populations of stars \citep[e.g.][]{am09,tk10,dh10,sh11b}. It can now be applied to the hundreds of solar-type stars for which \textit{Kepler} has detected oscillations \citep{wc11a}.
\section{Data analysis and stellar properties determination}\label{data}
\begin{figure*}
\plotone{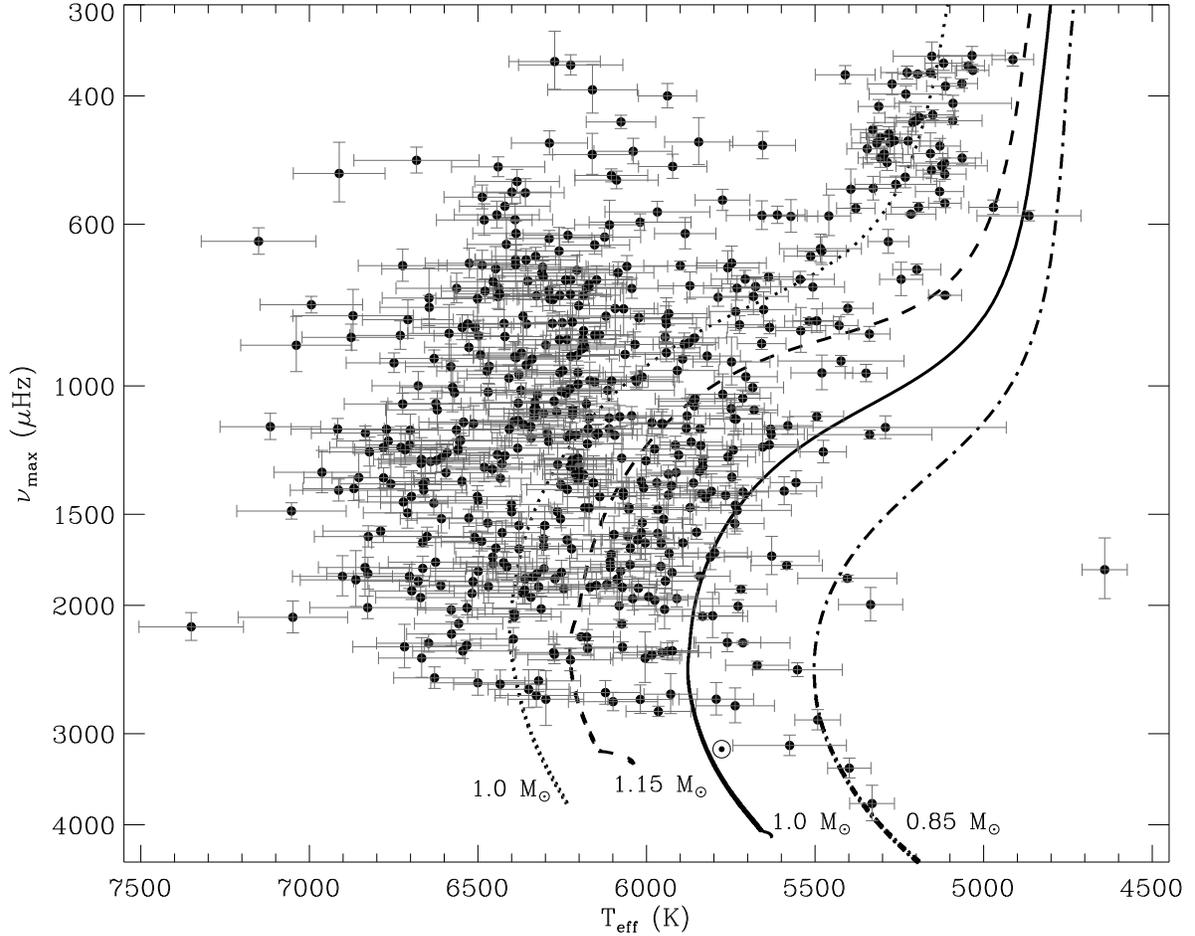}
\caption{$T_\mathrm{eff}$-$\nu_\mathrm{max}$ diagram for the complete sample of targets with detected oscillations. Evolutionary tracks are also plotted: at solar ([Fe/H]=0.0) metallicity for 0.85~M$_\odot$ (dash-dotted line),  1.0~M$_\odot$ (solid line) and  1.15~M$_\odot$ (dashed line), and at sub-solar metallicity ([Fe/H]=-0.5) for 1.0~M$_\odot$ (dotted line). The Sun is marked with its usual symbol.}
\label{astdiag}
\end{figure*}
We examined asteroseismic results for 2000 solar-type stars observed by \emph{Kepler} during the first seven months of science operations, with oscillations detected in more than 500 of them \citep{wc11a,gv11}. Each star was observed for one month in short-cadence mode \citep[58.85\,s sampling; see][]{gill10b}. Time series were prepared for asteroseismic analysis as described by \citet{rg11}, using procedures that work on the raw light curves. The prepared light curves were analyzed by different teams, who detected and extracted the basic properties of the solar-like oscillations. The methods of extraction and descriptions of each pipeline may be found in \citet{wc10,dh09,ma09,sh10,sm10,tc10}, and \citet{ck11}.

Effective temperatures have been derived via the InfraRed Flux Method (IRFM) described by \citet{lc10}, using the available Sloan and 2MASS photometry compiled for the targets in the Kepler Input Catalog \citep[KIC;][]{tb11}.

Figure~\ref{astdiag} shows a $T_\mathrm{eff}$-$\nu_\mathrm{max}$ diagram constructed from the consolidated results of different pipelines for all the targets with oscillations measured by at least two of them \citep[as described by][]{gv11}. Stars with $\nu_{\mathrm{max}}$ below 350\,$\rm \mu Hz$ are beyond the subgiant phase of evolution and have been omitted from our sample. We have overplotted evolutionary tracks for different masses and metallicities calculated with the GARching STellar Evolution Code \citep[GARSTEC,][]{ws08}, scaled from the \citet{gs98} solar mixture considering a helium enrichment ratio of $\Delta\,Y / \Delta\,Z = 1.8$ \citep{lc07} and not including diffusion of helium and heavy elements (which is why the solar metallicity 1-M$_\odot$ track does not overlap the position of the Sun).

It is already clear in Fig.~\ref{astdiag} that a number of oscillating stars are located on or very close to the evolutionary tracks and thus are potentially 1-M$_\odot$ stars. Interestingly, the stars are concentrated in two main regions, with an apparent gap between them (around $T_\mathrm{eff}\sim$5300 K and $\nu_{\mathrm{max}}\sim$600\,$\rm \mu Hz$). This feature suggests a rapid evolutionary stage where few stars are found, which could be identified as the transition between the main-sequence turn-off and the base of the red giant branch. It has also been suggested that an increase in the surface magnetic activity due to evolution can account for the lack of detections in this region, since activity depresses the oscillation amplitudes \citep{rg85}. The concentration of stars above the gap (around $T_\mathrm{eff}\sim$5200 K and $\nu_{\mathrm{max}}\sim$400\,$\rm \mu Hz$) further supports the latter idea.
\section{Results}\label{reslt}
\begin{figure*}
\plotone{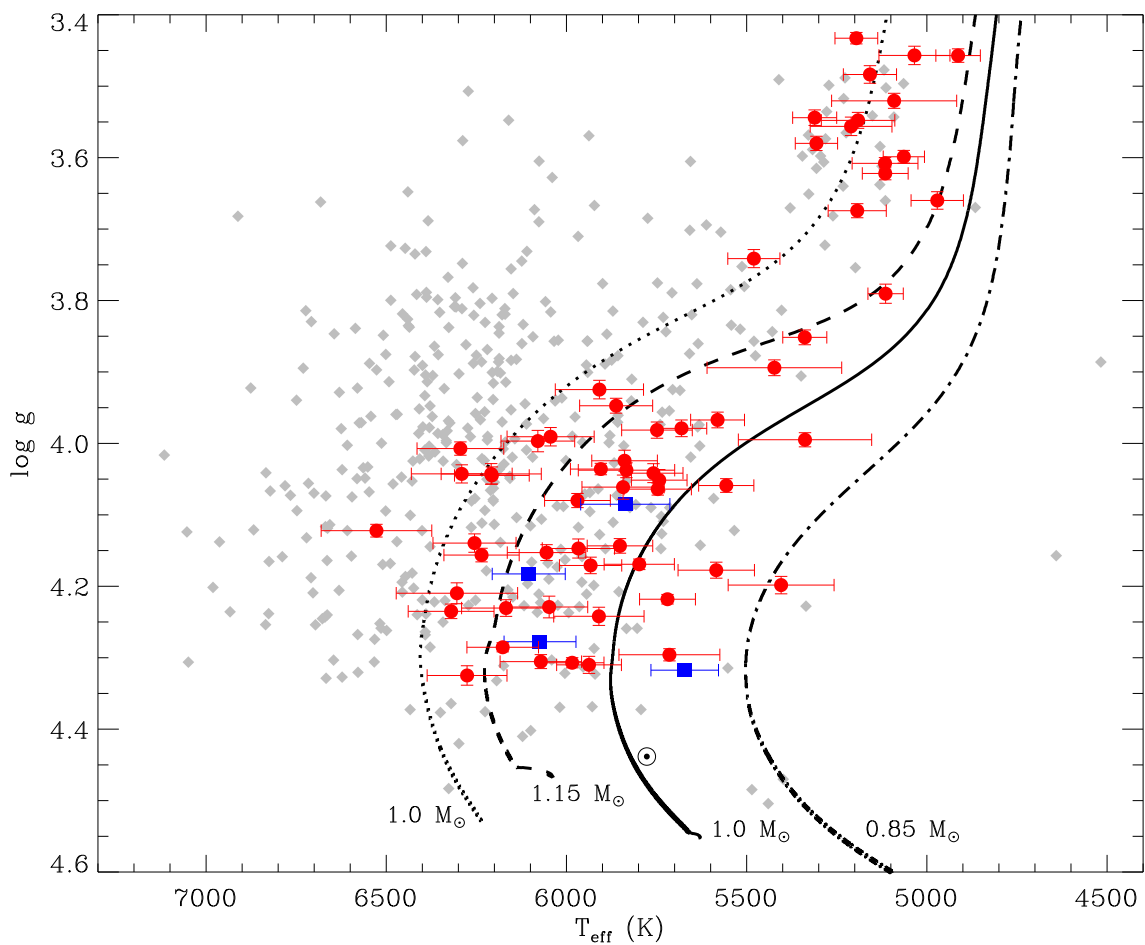}
\caption{Position in the  $\log\,g$-$T_\mathrm{eff}$ plane of the targets, where $\log\,g$ was obtained from the scaling relations. Stars with masses determined to be 1-M$_\odot$ $\pm$ 15\% are plotted in red circles, and stars with oscillations detected that do not match the criterion are plotted as gray diamonds (without error bars to reduce clutter). Blue squares are stars with available spectroscopic results. Stellar tracks and position of the Sun are plotted as in Fig.~\ref{astdiag}.}
\label{matches}
\end{figure*}
\begin{table*}[!ht]
\centering
\small
\begin{tabular}{|c|c|c|c|c|c|c|c|}
\tableline
Star & Mass (M$_\odot$)\tablenotemark{a} & $T_\mathrm{eff}$ IRFM (K) & $T_\mathrm{eff}$\tablenotemark{b} (K) & $\log\,g$\tablenotemark{a} & $\log\,g$\tablenotemark{b} & [Fe/H]\tablenotemark{b} & Source \\
\tableline
$1$ & $0.96\pm0.051$ & $5671\pm94$ & $5715\pm82$ & $4.32\pm0.007$ & $4.31\pm0.14$ & $0.30\pm0.06$ & ARES\\
$2$ & $1.04\pm0.043$ & $6073\pm100$ & $6073\pm78$ & $4.28\pm0.006$ & $4.38\pm0.12$ & $-0.10\pm0.06$ & ARES\\
$3$ & $1.07\pm0.051$ & $6105\pm101$ & $5940\pm70$ & $4.18\pm0.006$ & $4.21\pm0.08$ & $-0.10\pm0.07$ & VWA\\
$4$ & $1.07\pm0.062$ & $5836\pm124$ & $5870\pm70$ & $4.09\pm0.008$ & $4.07\pm0.08$ & $-0.01\pm0.07$ & VWA\\
\tableline
\end{tabular}
\caption{Stellar parameters of the four targets with available spectroscopic observations.}
\label{tab_ch}
\tablenotetext{a}{From the direct method}
\tablenotetext{b}{From spectra}
\end{table*}

\begin{figure*}
\plotone{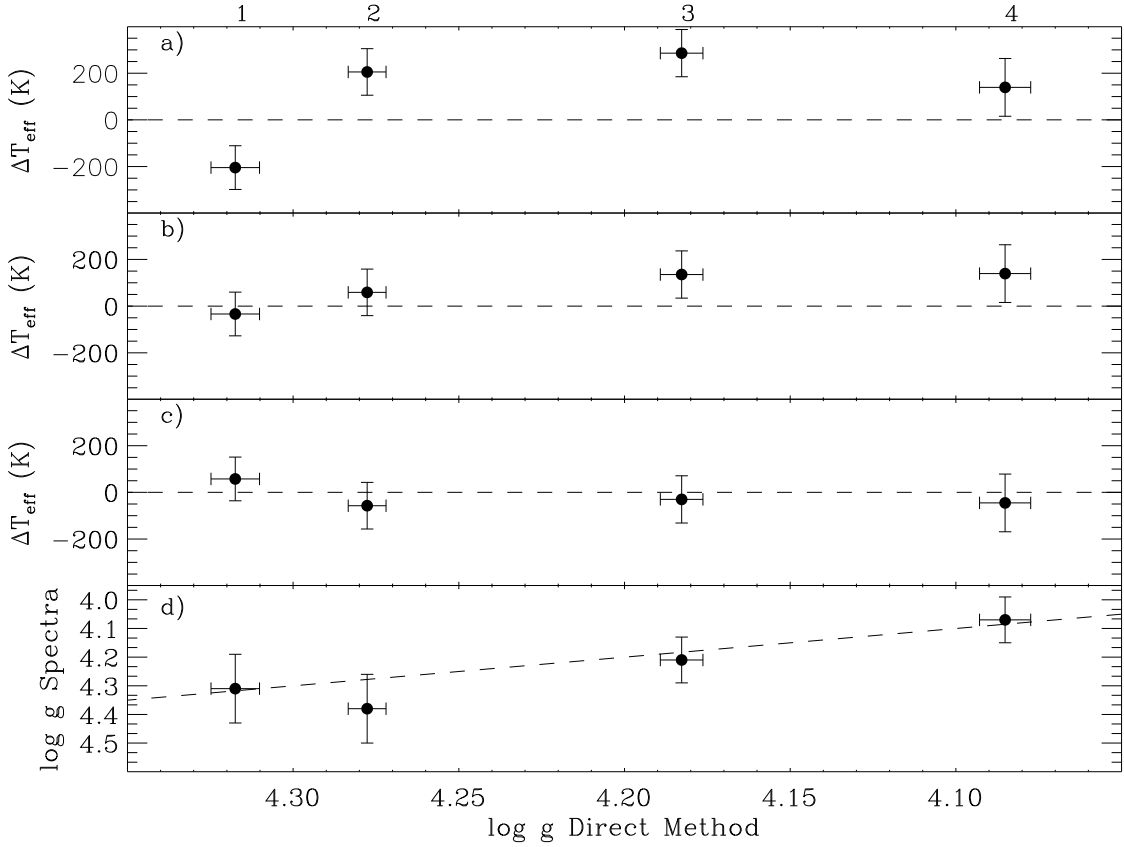}
\caption{The first three panels show the $T_\mathrm{eff}$ difference (IRFM-model) for the four targets marked in Fig.~\ref{matches} and: \textbf{a)} the 1-M$_\odot$ track at solar metallicity, \textbf{b)} a 1-M$_\odot$ track at the spectroscopic metallicity for that star, \textbf{c)} a track of the mass determined by the direct method and the spectroscopic metallicity for that star. Panel \textbf{d)} shows the $\log\,g$ values obtained from spectroscopy compared to those determined with the direct method , with the dashed line indicating the one-to-one relation.}
\label{chem}
\end{figure*}
\begin{figure}
\plotone{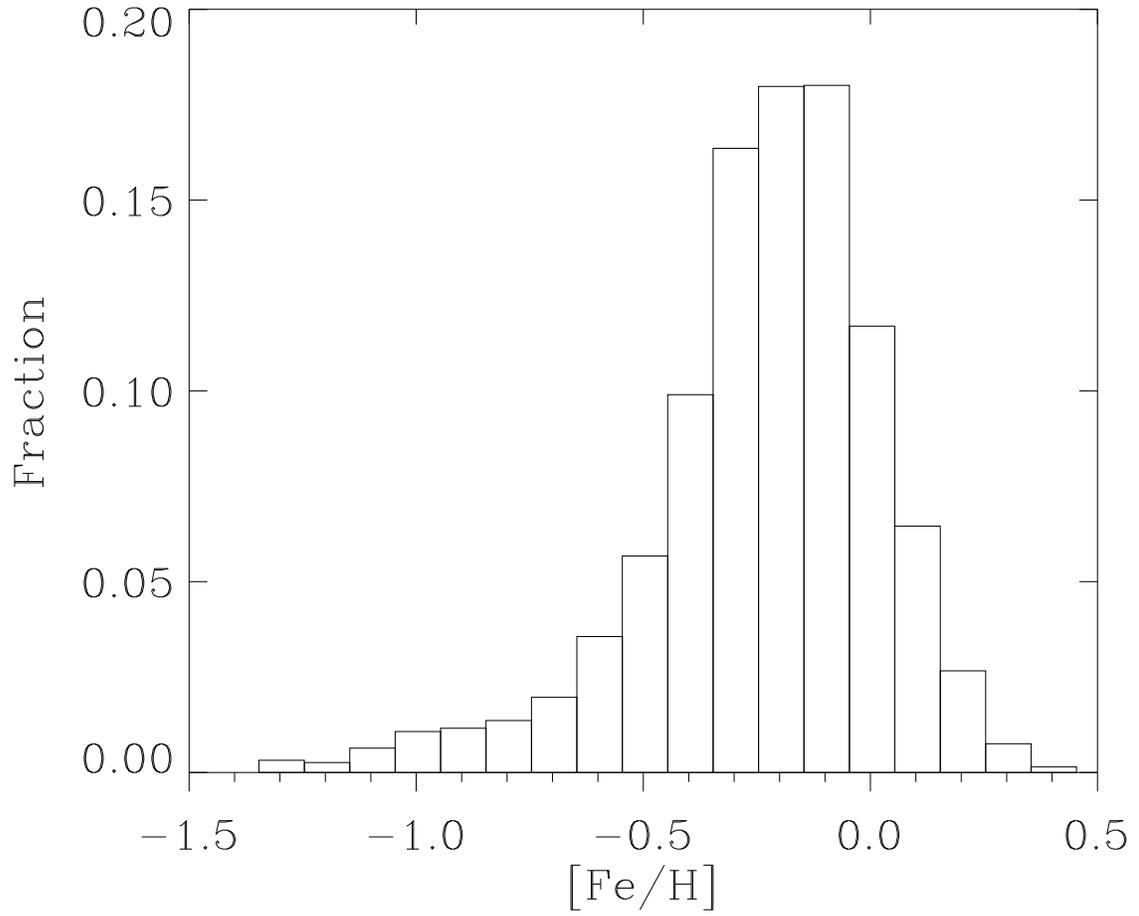}
\caption{Metallicity distribution in the \textit{Kepler} field as simulated by the SMAUG code, showing stars with a probability higher than 90\,\% to have oscillations detected after one month of observations.}
\label{histogram}
\end{figure}

We have applied the direct method to the sample in Fig.~\ref{astdiag} and determined $\log\,g$ values using the asteroseismic results extracted by the QML pipeline, which has been used as reference for comparison with other pipelines \citep{vr10,gv11}. To identify stars with masses similar to the Sun for constructing the evolutionary sequence, we selected targets that according to their determined masses and uncertainty, propagated from Eq.~\ref{eqn:mass}, had a probability larger than 68\,\% to be in the 0.85-1.15 M$_\odot$ range.

Figure~\ref{matches} shows these stars in the $\log\,g$-$T_\mathrm{eff}$ plane, where 72 targets are found to match the selection criterion. The distribution of matches in the observational plane is neatly contained by the stellar tracks, pointing towards an agreement between the asteroseismic mass values and evolutionary predictions. As in Fig.~\ref{astdiag}, a gap can be seen around $T_\mathrm{eff}\sim$5300 K and log(g)$\sim$3.7, again identified with the separation between the end of the main-sequence and the red giant phases.

In Fig.~\ref{matches} it is also clear that most of the matches fall on the hotter side of the solar-metallicity 1~M$_\odot$ track. There are two possible reasons for this. The first is that the amplitudes of the oscillations increase with increasing $L/M$ ratio \citep[or some variant of it, see][]{rs07,bm10,wc11b,kb95,kb11,dh11}, resulting in a detection bias in our ensemble that favors hotter (higher-than-solar-mass) main-sequence and subgiant stars. However, for our considered threshold, only 53\,\% of the matches have a determined mass larger than 1-M$_\odot$, suggesting a mass distribution barely skewed to higher values than our target mass. The second reason is a spread due to different chemical compositions of the stars because, for a given mass, a lower metallicity will imply a higher luminosity (and vice versa).

Keeping this in mind, we can infer from the position of the stellar tracks in Fig.~\ref{matches} that the spread is most likely due to differences in the chemical compositions of the stars. The effects of different masses and other physical processes such as microscopic diffusion (estimated by the separation between the 1-M$_\odot$ solar metallicity track and the position of the Sun) are not large enough to account for this spread. Unfortunately, we do not yet have spectroscopic determinations of metallicity for most of the stars we are studying, so the question arises: can we reliably identify stars with similar masses, or does the direct method provide incorrect mass determinations that are camouflaged by the unknown stellar parameters?

We can have a first glimpse at the answer by using spectroscopic results which are available for four of our targets (blue squares in Fig.~\ref{matches}). The stellar parameters have been determined using the ARES \citep{ss08} and VWA \citep{hb10} methods, where we selected the spectroscopic set of parameters with  $\log g$ value closer to the direct method determination. These are shown in Table~\ref{tab_ch}. It can be seen that our $T_\mathrm{eff}$ input values (IRFM) for these four stars are compatible with the spectroscopic results.

In Fig.~\ref{chem} we compare the results of the direct method for the four stars with evolutionary predictions by showing deviations in $T_\mathrm{eff}$ between our IRFM input parameter and stellar tracks at the $\log\,g$ value determined from the direct method. Figure~\ref{chem}a shows the $T_\mathrm{eff}$ difference between our targets and a 1-M$_\odot$ evolutionary track at solar metallicity, with large deviations arising from a combination of different metallicities and masses. The following panels show the results when we take these differences into account step by step. Firstly, Fig.~\ref{chem}b depicts the difference in $T_\mathrm{eff}$ between our targets and 1~M$_\odot$ tracks at the metallicity obtained from spectroscopy. Stars 1 and 2 are already compatible with the stellar tracks, which is encouraging as their masses determined with the direct method are consistent with 1-M$_\odot$ within their 1-$\sigma$ uncertainties. The other two stars have slightly higher-than-solar mass determinations. In Fig.~\ref{chem}c the difference is calculated with respect to evolutionary tracks of the metallicity measured by spectroscopy and mass determined from the direct method. It is clear that the agreement improves as we include the effects of mass and chemical composition, reinforcing the conclusion that the direct method predicts masses for field stars compatible with evolutionary tracks of the metallicities obtained from spectroscopy, in a model-independent way. Moreover, the $\log\,g$ values determined with the direct method are in agreement with the spectroscopic results, as shown in Fig.~\ref{chem}d.

To obtain some information on the expected metallicities for the rest of our targets, we simulated the stellar population in the \textit{Kepler} field using the Simple Model for Analytic Understanding of our Galaxy code \citep[SMAUG,][]{sb09a,sb09b}. The outcome of the simulation has been filtered to retain only synthetic stars that, according to their radius, effective temperature and visual magnitude, have a probability greater than 90\% for the \textit{Kepler} satellite to detect oscillations in one-month-long observations \citep{wc11b}. The resulting subset has been further restricted to stars with $T_\mathrm{eff} \leq$ 7000 K, $\log\,g$ $\geq$ 3.4 and $m_\mathrm{v} \leq$ 12.5. The metallicity distribution of that population is shown in Fig.~\ref{histogram} and peaks at a sub-solar value, with the bulk of stars contained in a range that makes them clearly compatible with the position of our matches and stellar tracks in Fig.~\ref{matches}.
\section{Discussion}\label{disc}
The sequence of field stars identified in the previous section shows very good agreement with the expected position of 1-M$_\odot$ stars in the $\log\,g$-$T_\mathrm{eff}$ plane. For those targets with metallicity measurements available, the masses and radii determined with the direct method are fully compatible with evolutionary tracks and spectroscopic $\log\,g$ determinations. Nevertheless, we must keep in mind that uncertainties in the observables and the scaling relations can affect our derived quantities.

From Eqs.~\ref{eqn:mass} and \ref{eqn:rad} it is clear that the mass and radius determinations are mostly sensitive to the uncertainties arising from the asteroseismic parameters. Although the errors in effective temperature determinations cannot be neglected, the fractional uncertainties in $\nu_{\mathrm{max}}$ are currently up to four times larger than those in $T_\mathrm{eff}$ and $\Delta\nu$. Therefore, the uncertainties in the frequency of maximum oscillation power currently account for most of the error budget.

Masses and radii determined from the direct method have an intrinsic uncertainty associated with two effects: deviations from the scaling relations and unknown stellar properties that can change the pulsation characteristics of the star. The scaling relation between the large frequency separation and the mean stellar density has been shown by theoretical models to be quite robust \citep{ds09a,tw11}. It is more complicated to estimate the uncertainties arising from the $\nu_{\mathrm{max}}$ scaling, since it relies on the relation between $\nu_{\mathrm{max}}$ and the acoustic cut-off frequency under the assumption of an isothermal atmosphere \citep[e.g.][]{tb91}. Although a partial theoretical basis for this relation has only recently been developed \citep{kbe11}, comparisons under this assumption with stellar models predict that it holds within a few percent for cool models \citep[close to solar-mass, see][]{ds09a}. Observational results for stars whose parameters are accurately known confirm this result \citep{bk03}. Realistic model atmospheres could help to assess its deviations, as could comparisons with direct observational methods to measure stellar properties, such as long-baseline interferometry.

The other main source of uncertainties is related to unknown stellar parameters, such as chemical composition. It is possible to complement the direct method with our knowledge of stellar evolution and match the seismic observables to their expected values from previously calculated evolutionary models. This so-called \textit{grid-based method} has been shown by simulations to perform as well as the direct method for constraining the radius \citep{ds09b,sb10,tk10}. When accurate metallicity measurements are available, the grid-based method can reduce the uncertainty in the mass determination \citep{ng11}. A combination of both methods will allow us to increase the precision and make the results even more robust. Independent tests are currently being developed to assess their uncertainties \citep[see, for example, Sect. 3.3 in ][]{wc11a}. Another validation of our results can be made by comparing the $\log\,g$ values determined with the direct method to the parallaxes that will be obtained from the \textit{Kepler} and Gaia missions in the future, given that a variation in mass of 10\,\% translate into a change in $\log g$ of approximately 0.04 dex \citep{lc11}.

In this work, we constructed the evolutionary sequence by applying the selection criterion to the parameters extracted by one particular pipeline. Different pipelines produced similar results in terms of the distribution of the matches in the HR diagram and the spread around the evolutionary tracks (see Fig.~\ref{matches}). A natural step forward would be to combine the results of several pipelines to restrict further the matches and make the results even more robust. However, there is still some variation between the extraction methods in the determination of $\nu_{\mathrm{max}}$. The reason probably lies in the low signal-to-noise ratio of some of the observations and the relatively short time-span of the data currently available \citep{gv11}.

Finally, our selection criterion can be arbitrarily modified to accept, for instance, stars with masses within 10\,\% of 1.0~M$_\odot$. In such case only 20 targets satisfy the criterion, while increasing the threshold to 20\,\% results in 120 matching stars. We have kept the 15\,\% threshold as a good compromise between a required high precision and the uncertainties described in the previous paragraphs.
\section{Conclusions and further development}\label{concl}
Constructing empirical evolutionary sequences from asteroseismic data offers the exciting possibility of performing differential analysis on field stars of similar masses, allowing us to test stellar properties very precisely through different evolutionary phases. Scaling relations using global asteroseismic parameters and $T_\mathrm{eff}$ measurements allowed us to determine masses of stars in a model-independent way. The position of our matches in Fig.~\ref{matches} suggests that we have successfully identified an evolutionary sequence of field stars with masses very close to 1~M$_\odot$. The results are encouraging and for the first time we can construct such a sequence without parallax information, spectroscopic $\log\,g$, or masses estimated from evolutionary tracks. Longer time series and ground-based follow-up spectroscopy will further enhance the capabilities of this technique, allowing us to reach a higher level of precision.

The results from Sect.~\ref{reslt} indicate that our mass and radius determinations are very robust. Using these results and the input effective temperature, together with a grid of stellar models, it should be possible to derive the expected metallicity distribution of the targets. This distribution will depend on the input parameters used to calculate the grid, most notably the selected solar abundances, $\Delta\,Y / \Delta\,Z$ value, $\alpha_\mathrm{MLT}$ parameter, and the inclusion (or not) of mixing processes such as diffusion. Future comparisons of this metallicity distribution with accurate chemical composition determinations could help to constrain the input parameters of the grid models that produce the largest variation in the resulting metallicities.

When suitable analogues are found in different evolutionary stages and their individual frequencies determined, a differential analysis can be performed to better constrain the interior physics of stars throughout their evolution. If targets with masses higher than solar are chosen instead, we could differentially study the effects of mixing processes in their interiors and constrain processes that still rely mostly on empirical calibrations, such as convective core evolution and overshooting \citep[e.g.][]{pdm10,vsa11}. For example, by selecting a sequence of stars of very similar mass it could be possible to produce a precise relative age calibration.

Asteroseismology has provided us with a method to disentangle stars still ascending the red giant branch from those already burning helium in their cores \citep{tbe11}. Our work can be extended to identify evolutionary sequences from the main-sequence to the red clump. Differential analysis of these targets can provide further constraints on the main-sequence physical processes and shed light on the progenitors of horizontal branch stars.

In the near future, it will be interesting to construct evolutionary sequences and select twins of very well-studied main-sequence stars, both seismically and in terms of their stellar parameters. From the ensemble it is already possible to identify sequences for $\alpha$\,Cen A and B, although for the latter case the number of analogues is much smaller due to our bias towards higher amplitudes than the Sun. Some twins of $\alpha$\,Cen A are already present in the ensemble, and longer time-series will allow us to expand the search to lower temperatures and find twins of our Sun and $\alpha$\,Cen B. The potential of combining asteroseismic data with spectroscopic determinations for solar twins has only recently begun to be exploited \citep{mb11}.

\acknowledgements 
Funding for this Discovery mission is provided by NASA's Science Mission Directorate. The authors wish to thank the entire \emph{Kepler} team, without whom these results would not be possible.  We also thank all funding councils and agencies that have supported the activities of KASC Working Group\,1. We are also grateful for support from the International Space Science Institute (ISSI).

\end{document}